\documentstyle[twocolumn,aps,psfig]{revtex}
\begin{document}
\draft
\title{Cooperative Transport of Brownian Particles}
\author{Imre Der\'enyi$^*$ and Tam\'as Vicsek$^\dagger$}
\address{Department of Atomic Physics,
E\"otv\"os University, Budapest, Puskin u 5-7, 1088 Hungary \\
  {\tt $^*$derenyi@hercules.elte.hu} {\rm ~~and~~}
  {\tt $^\dagger$h845vic@ella.hu} }


\address{~\\ \large \rm Physical Review Letters {\bf 75}, 374-377 (1995)}

\address{
\begin{minipage}{5.55in}
\begin{abstract}\hskip 0.15in
We consider the collective motion of finite-sized, overdamped Brownian
particles (e.g., motor proteins) in a periodic potential.  Simulations
of our model have revealed a number of novel cooperative transport
phenomena, including (i) the reversal of direction of the net current as
the particle density is increased and (ii) a very strong and complex
dependence of the average velocity on both the size and the average
distance of the particles.
\end{abstract}
\vskip -0.5in $ $
\end{minipage}
}

\maketitle

The most common and best known transport phenomena occur in systems in
which there exist macroscopic driving forces or gradients of potentials
of various origin (typically due to external fields or concentration
gradients).  However, recent interesting theoretical and experimental
studies have shown that non-equilibrium dissipative processes in
structures possessing vectorial symmetry can induce macroscopic motion
on the basis of purely microscopic effects
\cite{magn,ajd_pros,ast_bier,doer,mill_dyk,mill,bart,peskin,pr_aj,aj_pr}.

This newly suggested mechanism is expected to be essential for
biological transport processes such as the operation of molecular
combustion motors or the contraction of muscle tissues.  In these cases
Brownian particles (myosin, kinesin and dyenin) convert the energy of
ATP molecules into mechanical work while moving along periodic
structures (myosin along actin filaments, kinesin and dyenin along
microtubules) \cite{ash,svo,fin}.
A transport mechanism of this kind has also been
experimentally demonstrated in simple physical systems
\cite{r_aj_pr,fau}.

So far the models for the transport of Brownian particles in periodic
structures have been based on the description of the motion of one
single particle, but in real systems one can rarely find this situation.
Experimental evidence shows that several motor proteins can carry one
larger molecule, and a large number of free motors can move along the
same microtubule \cite{ash}.
Furthermore, in separation processes a large number
of particles are moving in the same medium \cite{r_aj_pr}.

Therefore, we propose a simple one-dimensional model {\it via many
interacting Brownian particles} moving with overdamped dynamics in a
periodic potential.  According to our computer simulations the model
displays a number of novel cooperative phenomena. i) First we show that
for a range of frequencies of a {\it periodic} external driving force
the {\it average velocity} $v$ of the particles {\it changes its
direction} as the number density of the particles is increased. ii) In
addition, $v$ has a sensitive dependence on the {\it size} of the
migrating particles.  This effect is demonstrated for two kinds (periodic
and constant) of driving forces.  In the last part of the paper we
present analytical results indicating that in the case of {\it constant}
driving force and nearly zero distance between the particles the
dependence of the velocity on the particle size becomes {\it extremely
complex} (non-differentiable).

\begin{figure}
\vskip -0.1in
\centerline{\psfig{figure=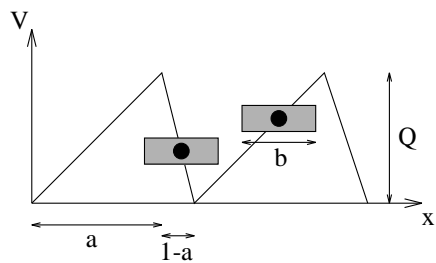,height=1.5in}}
\caption{
Schematic picture of the system we consider showing two particles
with size $b$ subject to the sawtooth shaped periodic potential $V(x)$.
The period of the potential is $p=1$, where the lengths of the slopes are
$\lambda_1=a$ and $\lambda_2=1-a$.
The potential difference between the top and the bottom is $Q$.
}\label{fsaw}\end{figure}

The motion of a single particle (in the absence of other particles) is
described by the Langevin equation

\begin{equation}
\dot x_j = f(x_j)+\xi_j(t)+F_j(t), \qquad j=1,...,N ,
\label{eq_lang}
\end{equation}

\noindent where $N$ is the number of particles, $x_j$ denotes the
position of the center of mass of the $j$th particle, $f(x)\equiv
-\partial_xV(x)$ is a force field due to the sawtooth shaped periodic
potential $V(x)$, $\xi_j(t)$ is Gaussian white noise with the
autocorrelation function
$\bigl<\xi_j(t)\xi_i(t')\bigr>=2kT\delta_{j,i}\delta(t-t')$, and
$F_j(t)$ is a ``driving force" with zero time average, which may be
stochastic.  Since in most of the experimental situations we can suppose
that the interaction between two particles can be well approximated with
a hard core repulsion we assumed that the {\it particles are hard rods}
(see Fig.\ \ref{fsaw}). The hard core interaction means that during the
motion the particles are not allowed to overlap (a particle does not
continue to move in its original direction if it touches another one).
This rule complements eq. (\ref{eq_lang}).
All particles have the same size $b$, while the period of the potential
is $p=1$.  The size of the system or in other words the number of the
periods is $L$.  We have applied periodic boundary conditions. The
position of the particles were updated sequentially (one after
another, from left to right) using the finite difference version of
(\ref{eq_lang}).  We have checked other types of
updatings (random, from right to left), with no change in the results.
 $N$ and $L$ go to infinity, while $L/N$ remains
finite, but usually $N\approx 20$ is large enough.  During the
integration of (\ref{eq_lang}) $\delta t$ was typically equal to 0.0001.
Most of the runs required several days on a fast IBM RISC 6000/375
workstation.

Our model is one dimensional because the macromolecules serving as
highways in biological transport can be assumed to be linear
representing well defined tracks.  Thus, due to the hard core
interaction we also exclude the possibility of "passing".  In higher
dimensions (where the particles can get around each other) further
effects are expected to take place.  In addition to the case of periodic
driving force in the second part of the paper we shall also consider the
case of constant driving force, because the latter case is i)
conceptually simpler, thus, allows more direct interpretation of the
simulational results and the analytic treatment of some limiting cases
and ii) a zero-mean signal can always be constructed as an alternating
($+F$ and $-F$) piecewise constant signal.

\begin{figure}
\vskip -0.1in
\centerline{\psfig{figure=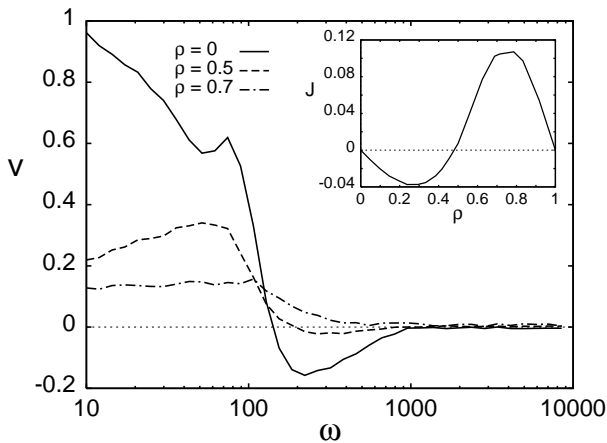,height=2.5in}}
\caption{
The plot of the average velocity $v$ as a function of the
average frequency  $\omega$
of the sinusoidal driving forces  for three different values of
the average covering $\rho\equiv bN/L$.
The inset demonstrates the reversal of the particle current $J\equiv
vN/L$ as a function of the average covering $\rho$, for $\omega=175$.
($Q=4$, $a=0.8$, $b=0.5$, $T=1$ and the amplitude of the driving forces $A=32$.)
}\label{fv_om_sin}\end{figure}

Normally, one single particle moves in the direction corresponding to
the smaller uphill slope of the potential.  However, there is a range of
the parameters of the {\it periodic} driving force for which the particle
migrates into the opposite direction \cite{doer,mill_dyk,mill,bart}.
In this regime we have found that {\it gradual addition
of particles} into the system {\it results in the change of their
average velocity} back to
the ``normal" direction.  We have tested this result for several different
cases (including driving forces periodic in time \cite{bart} and distributed
according to ``kangaroo" statistics \cite{doer}) and we have found that this
change of the current's direction
is a universal property of the collective motion in our model.
Fig.\ \ref{fv_om_sin} shows a simple
example, where the driving forces are $F_j(t)=A \sin(\omega_jt)$ and the
$\omega_j$ values are chosen randomly around a fixed value
$\omega$ with a dispersion of several percentage of $\omega$  (to
avoid synchronization).  The plot shows the average
velocity as a function of $\omega$, for various values of the average
covering defined as $\rho\equiv bN/L$  ($0<\rho<1$).
In the inset we have plotted
the fundamental diagram: the particle current $J\equiv vN/L$ as a
function of the average covering for $\omega=175$.

Another interesting feature is observed if the average distance
between two neighbouring particles is fixed ($d\equiv L/N-b=const$)
and we are changing the size of the particles.

Before describing our results we mention that it is easy to show that a
system of length $L$ consisting of $N$ particles of size $k+b$ ($0\leq b<1$,
$k=1,2,...$) is equivalent to a system of length $L-kN$ consisting of
$N$ particles of size $b$.  Obviously, this kind of transformation has
no effect on the motion of particles, therefore, it is enough to
consider particles with sizes less than one.  In other words, any
quantity is a periodic function of the size of the particles with period 1
{\it i.e.} with period equal to the period of the underlying
potential.

\begin{figure}
\vskip -0.1in
\centerline{\psfig{figure=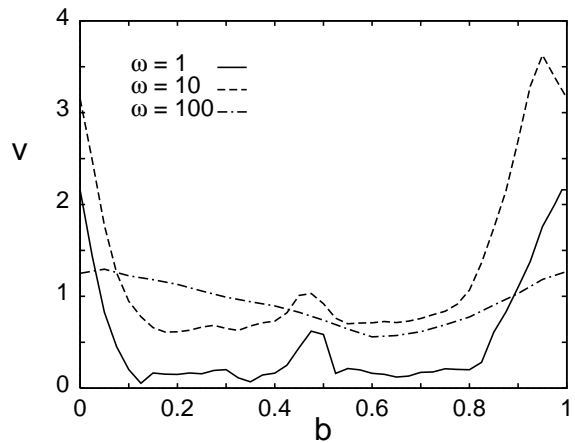,height=2.5in}}
\vskip -0.02in
\caption{
The plot of the average velocity $v$ as a function of the size of the
particles $b$ for three different values of the frequency $\omega$
of the sinusoidal driving forces.
The average distance between two neighbouring particles is $d=0.5$.
($Q=4$, $a=0.8$, $T=1$ and $A=32$.)
}\label{fv_b_sin}\end{figure}

Fig.\ \ref{fv_b_sin} shows the average velocity as a function
of the size of the
particles in the above mentioned case with sinusoidal driving forces for
various values of $\omega$.  The velocity has very drastic changes.  A
large peak can be observed for $b$ somewhat smaller than 1, and a
smaller peak for $b$ somewhat smaller than 1/2.  In most of the other
cases we have studied, a large peak is observed just before $b$
reaches 1 or for $b$ a bit larger 0 (or equivalently larger than 1), and
a minimum (valley) on the opposite side of this integer value.
This structure is repeated around 1/2, but on a smaller scale.
Sometimes this structure can be observed around 1/3 and 2/3.

Investigating the origin of this strange behavior of the particle size
dependence on the average velocity we examine the simplest case when the
driving force is {\it stationary}: $F_j(t)=F$ and smaller than the
uphill gradient of the potential.

\begin{figure}
\centerline{\psfig{figure=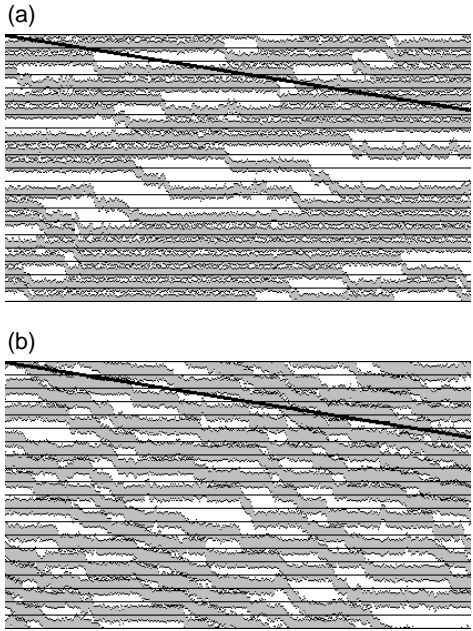,height=3.3in}}
\bigskip
\caption{
Motion of the particles in the space-time domain. The time increases
from left to right and the particles are moving downwards
under the influence of the stationary driving force $F$.
Horizontal lines represent the bottom of the potential-valleys, and
the wide slanted line represents the average motion of a single
noninteracting particle.
(a) Shows 15 particles with size $b=0.833$. A vacancy type current
can be observed, as a consequence of the hindering effect of particles.
The average velocity $v$ is smaller than the average velocity
of one single particle.
(b) Shows 12 particles with size $b=1.166$.
There are no jams, and the density waves show that the particles assist
each other in jumping over to the next valley.
$v$ is larger than the velocity
of a single particle.
The average distance between particles is $d=0.5$ in both cases.
}\label{ftr}\end{figure}

Let us consider the case when the size of the particles is somewhat less
than 1 and there are two particles in the neighboring valleys of the
potential.  Then the second particle is not able to jump further ahead
until the first one jumps away.  So the first one hinders the second
one.  Thus, the average velocity is smaller than the velocity of a
single particle.  Fig.\ \ref{ftr}a shows this situation for 15 particles.
A {\it vacancy type} current can be observed, as a consequence of the traffic
jams arising from the hindering of particles.  This phenomenon is
also related to jams
common in one dimensional driven diffusive systems and traffic
models \cite{evans}. If the size of the
particles is a bit larger than 1 and there are also two particles in the
neighboring valleys, both of them cannot be in the minimum in the same
time, therefore, the first one has a larger chance to jump further.  In
this case the second one indirectly ``pushes" the first one.  (But the
first one also hinders the second one.)  Thus, in spite of the hindering
effects the average velocity can be larger than the velocity of a single
particle.  This situation can be seen in Fig.\ \ref{ftr}b for 12 particles.
There are no jams and the density waves show that the particles help
each other to jump through to the next valley. In case of slowly
alternating external forces these effects (hindering and pushing) are
expected to influence the net transport.

\begin{figure}
\centerline{\psfig{figure=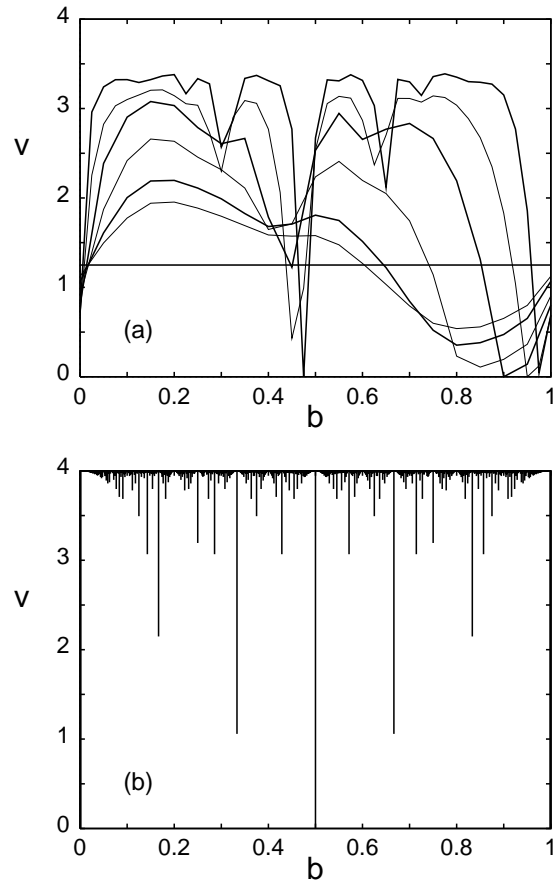,height=4.8in}}
\caption{
The average velocity $v$ as a function of the size of the
particles $b$, when the driving force is stationary with $F=4$.
(a) The plot for different values of the average distance between two
neighbouring particles: $d=\infty$ (one single particle, the horizontal line)
and $d=0.6$, 0.4, 0.2, 0.1, 0.05, 0.025.
(b) The plot in the limit when the average distance between two
particles goes to zero. This discontinuous function has sharp minima
for $b$ rational and a value equal to $F$ if $b$ is irrational.
($Q=4$, $a=0.2$ and $T=1$.)
}\label{fv_b_stat}\end{figure}

Fig.\ \ref{fv_b_stat}a shows the average velocity as a function of the size
of the particles in this stationary case, for various values of the average
distance $d\equiv L/N-b$ between two neighboring particles.  When $d$ is
infinity, the velocity is independent of the size of the particles, and
identical to the velocity of a single particle.  Decreasing $d$ a velocity-peak
starts to develop for $d$ just larger than $b=0$, and a valley appears
for $b$ close to, but smaller than $b=1$.  This was explained in the
previous paragraphs.  As $d$ is further decreased, another peak appears
beyond $b=1/2$ and also a valley before $b=1/2$.  This can also be
explained in the above mentioned manner, taking into consideration that
two particles can sit in the same potential-valley if $b\approx 1/2$,
and we can handle them as one particle with size $b\approx 1$.
Decreasing the average distance further, valleys and peaks appear before
and after $b=1/3$ (3 particles in 1 potential-valley), $b=2/3$ (3 particles
in 2 potential-valley), and so on at (almost) any rational value of $b$.

If the sum of the average distance and the size of the particles is a
rational value, i.e., $b+d=n/m$, we can say that the structure is {\it
commensurate}. For $d\ll 1$ the particles are distributed evenly and
$m$ particles can be found in $n$ potential-valleys. The minimum of the
potential energy of the system is realized if every $m$th particle is
sitting in the bottom of the potential valleys. Then, for $F=0$ each
particle has to jump a distance $1/m$ to reach the next minimum energy
state of the system.
Simple algebra shows that such a system (in which $N$ particles
are playing the role of a single particle) can also be described in
terms of a modified sawtooth potential with a period $p'=1/m$,
where the lengths of the slopes are
$\lambda_1'=\{ma\}/m$ and $\lambda_2'=\{m(1-a)\}/m$.
The potential difference between the top and the bottom states is
$N\cdot Q'$, where

\begin{equation}
Q'=Q{\{ma\}\{m(1-a)\}\over ma\cdot m(1-a)}.
\end{equation}

\noindent The notation $\{...\}$ means the fractional part
of the value between the braces.

Thus, in the presence of the driving force $F$ we can calculate the
average velocity as the velocity of a single particle using the formula
derived by Magnasco \cite{magn} with parameters
$Q'$, $\lambda_1'$, $\lambda_2'$, $F'=F$ and $T'=T/N$ ($T'\to 0$ for
$N\to \infty$).

However, if the structure is incommensurate and the average distance is
small, the corresponding modified potential of the whole system is
almost flat
and the system has a continuous translation symmetry.  Therefore, the
particles can move with almost the maximum velocity $v_{max}=F$.

The modified potential is also flat (or almost flat), if the structure
is commensurate
but $ma$ is an integer number (or close to an integer number).
This is the reason why we can not see a valley before
1/5, 2/5, 3/5 and 4/5 on Fig.\ \ref{fv_b_stat}a as a consequence of $a=0.2$.

Correspondingly, decreasing the average distance between the particles
the minima of the valleys tend to the rational values.
The values of the minima go to the values calculated from Magnasco's formula,
and the width of the valleys goes to zero.
For the other cases the velocity goes to $v_{max}=F$.
In the limit when the average distance is zero, we get a
strange, discontinuous function with sharp minima for $b$
rational and a value equal to $F$ if $b$ is irrational
(Fig.\ \ref{fv_b_stat}b).

In conclusion, we have demonstrated that taking into account the
interaction of Brownian particles migrating via overdamped dynamics
along periodic structures results in a variety of novel cooperative
effects.  Among other possible applications, our results are expected to
be pertinent from the point of biological transport involving finite
density of protein molecules moving along substrates made of
macromolecules.  In particular, we have found a strong dependence of the
average current on the particle size for sizes close to the period of
the underlying potential.  If thermal ratchet type models represent
an adequate description of biological transport our latter result is
likely to be relevant in the understanding of the behavior
of such molecular motors as kinesin or dyenin since their
size and the period of the corresponding microtubules are
comparable (see, e.g., Ref. \cite{svo}).
Effects caused by the finite size of the transported objects
(e.g., other proteins, mitochondria, visualizing beads) represent
potential subjects for further studies.

\acknowledgments The authors are grateful to A. Ajdari and T.
Geszti for useful discussions. The present research was supported by the
Hungarian Research Grant No T4439.

\end{document}